\documentclass[acmsmall]{acmart}
\usepackage{multicol}
\usepackage{multirow}
\usepackage{tabularx}
\usepackage{makecell}
\usepackage{graphicx}
\usepackage{booktabs}

\AtBeginDocument{%
  }

\setcopyright{cc}
\setcctype{by-nc}
\acmJournal{PACMHCI}
\acmYear{2025} \acmVolume{9}
\acmNumber{7} \acmArticle{CSCW342} \acmMonth{11}
\acmDOI{10.1145/3757523}

\received{July 2024}
\received[revised]{December 2024}
\received[accepted]{March 2025}
\acmISBN{978-1-4503-XXXX-X/18/06}

\begin{document}

\title[When Technologies Are Not Enough]{When Technologies Are Not Enough: Understanding How Domestic Workers Employ (and Avoid) Online Technologies in Their Work Practices}

%%The Crescent Moon Effect: The Ambivalent Role of Technology in the Lives of Domestic Workers  
%How Technology Fails to Revolutionize Domestic Work
%\title{Dormant Technology in Domestic Wor}
%The Unmet Promise of Technology for Domestic Workers

\author{Mariana Fernandez-Espinosa}
\email{mferna23@nd.edu}
\orcid{0009-0004-1116-2002}

\author{Mariana Gonzalez-Bejar}
\email{mgonza32@alumni.nd.edu}
\orcid{0009-0002-7433-4145}

\author{Jacobo Wiesner}
\email{jwiesner@nd.edu}
\orcid{0009-0007-8679-4380}

\author{Diego Gomez-Zara}
\authornotemark[1]
\orcid{0000-0002-4609-6293}
\email{dgomezara@nd.edu}
\affiliation{%
  \institution{University of Notre Dame}
  \city{Notre Dame}
  \state{IN}
  \country{USA}
}

\renewcommand{\shortauthors}{Mariana Fernandez-Espinosa, Mariana Gonzalez-Bejar, Jacobo Wiesner, \& Diego Gomez-Zara}
\begin{abstract}
Although domestic work is often viewed as manual labor, it involves significant interaction with online technologies. However, the detailed exploration of how domestic workers use these technologies remains limited. This study examines the impact of online technologies on domestic workers' work practices, perceptions, and relationships with customers and employers. We interviewed 30 domestic workers residing in the United States, who provided examples that highlight the insufficient transformative role of current online technologies in their work. By conducting a thematic analysis, we characterize how they approach and avoid these digital tools at different stages of their work. Through these findings, we investigate the limitations of technology and identify challenges and opportunities that could inform the design of more suitable tools to improve the conditions of this marginalized group.
\end{abstract}





\maketitle

\section{Introduction}\label{sec:Intro}
The advancement of new sociotechnical systems has reshaped multiple aspects of labor and work practices. For example, crowd-sourcing platforms (e.g., UpWork, Amazon Mechanical Turk) and on-demand work platforms (e.g., Uber, Lyft, DoorDash, and GrubHub) have revolutionized workers' operations by employing algorithms and servers to coordinate their work \cite{scholz2017uberworked, healy2017should, Kittur2013, timko2018gig}. Moreover, automating technologies have transformed manufacturing and administrative tasks, reducing risks of accidents and making processes more efficient \cite{9765027}. Lastly, remote work technologies, including video conferencing and collaborative software such as Zoom and Slack, have redefined traditional office work, enabling employees to work and collaborate from virtually anywhere in the world \cite{cook2020global, mariani2023rise, dingel2020many, yang2022effects}. These significant changes in technologies have been the subject of extensive research, demonstrating their transformative impact on economies, productivity levels, and workforces \cite{ravenelle2021side, rosenblat2016algorithmic, woodcock2020algorithmic}. 

While these innovations have reshaped many forms of labor, not all work has benefited from such technological transformations. In particular, \textit{domestic work}---which encompasses jobs and tasks performed within households---has remained largely manual and invisible to these transformations \cite{international2021making}. Domestic workers perform multiple kinds of jobs at homes and private spaces, such as caring for children and seniors, maintaining clean and organized spaces, and providing comfort and companionship \cite{dol2023,wiego2023}. Despite their prominence and relevance in many societies, extensive research has pointed out the precarious work conditions \cite{Burnham_Theodore_2012}, imbalanced power dynamics \cite{cox1997invisible}, wage theft \cite{sternberg2019hidden}, lack of legal protections \cite{davis1993domestic}, and exposure to workplace abuses \cite{burnham2018living} they face. Notably, domestic workers around the world, with different demographic characteristics and educational levels, face systematic work challenges \cite{iom2023domesticworkers,jokela2018layers}. 

In particular, research on domestic work within the fields of Human-Computed Interaction (HCI) and Computer-Supported Cooperative Work (CSCW) is still in its early stages. Most of the work has addressed care work, including caregivers and nannies, which are a specific subset of domestic workers. Ming and colleagues \cite{mossberger2003virtual} demonstrated how inappropriate sociotechnical systems reinforce the invisibility of care workers. Similarly, Chen et al. \cite{10.1145/2441776.2441789} supported this notion by offering design guidelines to address the invisible emotional burdens of care work, aiming to develop more effective online technologies. Nevertheless, other forms of domestic work need further examination through a CSCW lens to inform the development of more effective socio-technical systems and appropriate design guidelines for domestic workers \cite{10.1145/2441776.2441789,gallienne1993alzheimer}. As such, examining how sociotechnical systems have reshaped domestic work could offer valuable insights to enhance current technologies, improve working conditions, and foster more equitable practices.

In this study, we investigate how domestic workers performing cleaning services have used online technologies across different stages of their tasks and activities. We delve into the motivations, perspectives, and specific reasons why domestic workers adopt online technologies in their daily routines. Furthermore, we analyze the impact of these technologies, exploring whether they act as a transformative force or play only a minimal role in domestic work. Are current online technologies positioned to drive a significant change in this sector? Can they potentially revolutionize how domestic work is performed while improving worker welfare? Such research grounded on the HCI and CSCW literature is fundamental to not only promoting domestic workers' empowerment but also improving their working conditions. Therefore, our research questions are:
\begin{itemize}
    \item \textbf{RQ1:} How do domestic workers perceive and use online technologies in their work practices, and what motivates their integration or avoidance?
    \item \textbf{RQ2:} What are the main outcomes of integrating online technologies into the work practices of domestic workers?
\end{itemize}

To address our research questions, we interviewed 30 house cleaners residing in the United States to explore how they employ online technologies in their daily activities and interactions with customers and managers. Drawing on studies of the digital divide \cite{wheeler2017opportunities,star1999layers}, invisible work \cite{,fox2017imagining}, and immigrant labor \cite{10.1145/3311957.3359503, glenn2010forced, hsiao2018technology}, we analyzed participants' experiences and testimonies. We uncover varying degrees and contexts of technology usage through thematic analysis \cite{braun2006using}, including some cases revealing underutilization or complete absence of these technologies. Based on these insights, we introduce the concept of the \textit{``Crescent Moon Effect''} to highlight the nuanced roles of technology perceived by domestic workers. As only certain parts of the moon can be seen despite its presence, we found that domestic workers selectively adopt certain technological features--often repurposing them in innovative and strategic ways--while disregarding others that fail to provide the necessary trust and information to enhance their work. As such, this analogy emphasizes that technological adoption by domestic workers is an individual, selective, fragmented, and more strategic process, rather than a uniform progression. Extending previous frameworks of digital divides, which often categorize technological adoption in levels or overlook the nuanced individuals' motivations to adopt or reject technologies, the concept of the Crescent Moon Effect accounts for the ways workers selectively engage with specific technologies based on trust and information needs. This perspective shows that while certain features can be easily adopted or repurposed, others remain ignored, showing the complex and situational nature of online technologies for these groups. The strategic application of online technologies highlights their limited role in truly transforming the broader conditions of their work conditions, which, in some cases, provide competitive advantages only to those capable of using them effectively. 

Our study contributions are twofold. First, we provide a holistic and unique perspective on the role of online technologies for domestic workers, revealing their interactions and perceptions regarding the adoption of technologies in their work. This exploration aims to guide the design of more suitable technologies tailored to their needs. Second, we provide a study of 30 semi-structured interviews using thematic analysis that reveals that domestic workers' hesitation to adopt technology is not due to a lack of skills or access. Instead, we found that many current online technologies do not meet their needs to be fully and effectively integrated into their work practices. The lack of inclusion and scope at the design stages of online technologies leads to this systematic and perpetual exclusion of these workers. These insights have real-world implications, potentially guiding the development of new systems that improve work conditions for domestic workers.

\section{Theoretical Background}
\label{section:background}
This section explores previous studies of domestic workers in multiple fields, including sociology, economics, and feminist studies. We then review CSCW and HCI contributions that offer insights and theoretical foundations to study how domestic workers adopt and employ online technologies. We examine digital divide theories, particularly the second and third levels, to understand the usage and impact of technologies on domestic workers in their work environment. Lastly, we review feminist studies and intersectionality in HCI that could shed light on the specific intersection between gender and domestic work.

\subsection{Domestic Workers}
Domestic workers provide essential household services such as caregiving, cleaning, and childcare within private residences \cite{DomesticWorkers, dol2023}. Their roles within private households often lead to isolation and limited opportunities to scrutinize their working conditions, with the diffuse and informal nature of their tasks \cite{macdonald2011shadow, coser1973servants, anderson2000doing} further amplifying the invisibility \cite{burnham2018living} and lack of recognition of their work \cite{cox2006servant, hondagneu1994regulating, dwhs2023}. Moreover, many of them are women as they have traditionally fit into roles characterized by caregiving and household management \cite{rodriguez2007latinas,roberts1997spiritual}.

House cleaning, a primary role among domestic workers, involves tasks such as maintaining cleanliness in households and businesses \cite{BLS2023, chicago2023, ilo2023}. Historically marked by subordination \cite{romero1987domestic}, this sector has evolved as many women have transitioned to entrepreneurship, establishing their own businesses and employing other women, reflecting their efforts to assert control over their work \cite{wrigley1991feminists, romero1988chicanas, romero2016maid}. These changes empower workers to improve labor conditions and challenge exploitation \cite{garcia1994maids, coble2020cleaning, lan2006global}. 

Domestic workers are disproportionately affected by poverty and face precarious conditions, including exclusion from labor protections, wage theft, sexual harassment, discrimination, exposure to toxic chemicals, and workplace injuries \cite{anderson2001just, Burnham_Theodore_2012, papadakaki2021migrant, jokela2018layers, romero2016maid}. Studies highlight their low wages, poor conditions, and power imbalances with managers and clients, which foster verbal and mental abuse \cite{sternberg2019hidden, arcand2020live, davis1993domestic, odeku2014overview, aclu2023, fernandez2021power}. They often are among the lowest-paid and most informal wage earners \cite{ilo2023minwage, browne2003intersection}. For instance, in 2021, house cleaners in the U.S. earned an average of only \$22,157 annually, compared to the national median income of \$52,112 \cite{FactSheetUSLabor, international2013domestic, pew2023}. Moreover, factors such as gender, race, class, and immigration status exacerbate their vulnerabilities \cite{duffy2020driven}. Patriarchal and capitalistic systems further marginalize women in this sector, viewing them as docile and naturally skilled labor \cite{khankhoje1984we}. Kaur-Gills \cite{kaur2022hyper} describes how these intersecting factors create ``hyper-precarities,'' restricting access to work welfare and intensifying disadvantages.

Immigrants, particularly low-skilled women, are overrepresented in the domestic work sector \cite{byrd2009dirty, gallotti2015migrant} since often it requires no formal qualifications or employment authorization \cite{cortes2023immigration}. This overrepresentation is further driven by cultural and societal stereotypes, which describe immigrants as hardworking, obedient, and naturally skilled in caregiving or cleaning, making them preferred candidates for employers \cite{labadie2008reflections}. They often perform these jobs and endure exploitative conditions influenced by gendered and racialized dynamics, as well as intermediary exploitation \cite{hondagneu2007domestica, anderson2015migrant, moors2008migrant}. For instance, the domestic workforce in the US is predominantly composed of women from racial and ethnic minority groups (95\%), with Latinx individuals representing 58.9\% of domestic workers, highlighting their disproportionate representation \cite{FactSheetUSLabor}. Domestic work serves as vital income for older individuals, with 40.2\% aged 50 or older, compared to 33.7\% in other occupations \cite{pew2023}. Immigrants also contribute significantly, participating at rates over 2.5 times higher than the general workforce\cite{usdol_domestic_workers_2021,gao1998}. 

%In the U.S., over 600,000 domestic workers are employed in private households, with 38\% working as house cleaners \cite{FactSheetUSLabor}. 

Extensive research has examined the socio-economic, cultural, and demographic factors shaping domestic work, but the role of technology remains underexplored. This study addresses how domestic workers adopt online technologies to improve their practices, focusing on the contributions and limitations of these technologies in enhancing their work conditions.

\subsection{Invisible Work, Low-wage, and Job Platforms}
Previous studies in CSCW and HCI have explored the reality that labor and low-wage workers face from different angles and lenses. We build on several of these studies to provide relevant theoretical insights to our study. First, CSCW researchers have extensively examined the concept of invisible work, defined as labor that remains unnoticed, unacknowledged, unvalued, or unregulated by employers, consumers, and legal systems \cite{hatton2017mechanisms, crain2016invisible}. Domestic work aligns with Hatton's invisible work framework \cite{hatton2017mechanisms}, which emphasizes how cultural, legal, and spatial dimensions limit workers' conditions and opportunities for improvement. Acknowledging domestic work as a form of invisible labor is essential for expanding its representation in CSCW research and addressing the challenges faced by these workers. Petterson et al. \cite{petterson2024networks} analyzed domestic work's private sphere, highlighting the invisible networks of care comprising people, technologies, spaces, and sociopolitical forces. Their work introduced the concept of ``lateral care'' to describe these digitally facilitated networks. Similarly, Ming \cite{ming2023data} investigated how data about care workers could inform the development of better sociotechnical systems, focusing on visibility and its implications for care work. 

Second, CSCW research has examined how low-wage workers use online technologies to enhance their skills, leverage their compensations, and address wage concerns. Dombrowski et al. \cite{LowPrecariousWorkers} examined the sociotechnical practices employed by US low-wage workers to combat ``wage theft,'' which involves illegal practices such as employee misclassification, minimum wage violations, and unauthorized pay deductions \cite{nelp2023wagetheft,kim2021wage}. Their study highlighted that workers identify payment discrepancies, document work hours, and organize collectively to address wage concerns. CSCW research has also explored employers' perspectives on recruiting low-wage workers. Social media, originally designed for personal connections, now plays a significant role in recruitment and job search for low-wage workers. Jiahong and Dillahunt \cite{lu2021uncovering} found that while employers rarely use social media to screen candidates directly, these platforms often reveal job seekers' readiness and commitment, influencing hiring decisions. In another study, Wheeler and Dillahunt \cite{wheeler2018navigating} observed that online platforms enhance the job search practices of low-resourced job seekers but fail to address issues such as limited digital literacy or access to devices. Similarly, Kingsley et al. \cite{kingsley2024your} emphasized the importance of clear and accurate job advertisements in helping low-income workers navigate the digital labor market successfully.

Lastly, CSCW research has explored disparities in access and outcomes across job recruitment and labor platforms. Low-income job seekers often favor platforms like Indeed and Facebook, which they find more accessible than professional tools like LinkedIn \cite{10.1145/3476065}. More specialized labor platforms such as `Care.com,' `Tidy,' and `Handy' connect domestic workers with clients for private home tasks but introduce risks like financial insecurity, job instability, and biases from rating systems and algorithmic management \cite{ticona2018beyond,collier2017labor}. While high-income job seekers employ diverse strategies to achieve better results in their searchers (e.g., networking, interview preparation), low-income job seekers do not count on the resources or digital skills to maximize their outcomes \cite{granovetter2018getting, wanberg2020job}. Although labor platforms have the potential to enhance job seekers' visibility and expand their reach to a broader range of opportunities, these platforms frequently lack resources to support diverse low-wage seekers. Prompting initiatives like `DreamGigs' have helped marginalized workers identify skill needs and connect with job opportunities \cite{wheeler2018navigating, kingsley2024your, 10.1145/3290605.3300808, dillahunt2018designing}. Moreover, nonprofits like the National Domestic Workers Alliance address these challenges through initiatives like ALIA, a job platform that offers portable benefits to domestic workers \cite{Chapter15}.

\subsection{Online Technologies and Domestic Work}
Online technologies have been increasingly adopted across various workforces to automate tasks, optimize workflows, and enhance coordination \cite{Kittur2013, baviskar2015comparative,katz2004emerging}. Digital platforms, in particular, have transformed sectors such as ride-sharing, food delivery, and knowledge-based freelancing by integrating algorithmic management and real-time communication to optimize labor coordination \cite{10.1145/2998181.2998327, fulker2024cooperation, DriverStake}.
Additionally, research has examined the emergence of digital marketplaces for job hiring within social media platforms such as Facebook groups and WhatsApp communities, highlighting how these online technologies serve as channels that facilitate new ways of finding employment \cite{jaimes2024costo,mim2022f, evans2018facebook, 10.1145/3411764.3445350}.

In the context of domestic work, prior research has primarily focused on digital labor platforms that facilitate hiring processes between workers and clients. Studies of these platforms have highlighted multiple challenges, including gender bias and precarious work conditions. For instance, Hannak et al. \cite{10.1145/2998181.2998327} found that on labor platforms such as ``TaskRabbit'' and ``Fiverr'', worker evaluations were significantly correlated with gender and race, potentially harming employment opportunities for domestic workers. Similarly, Lair et al. \cite{lair2018advertising} demonstrated that domestic work advertised on Craigslist often reflects precarious conditions, including low wages, broad job expectations, and little job security.

While existing research has extensively examined digital labor platforms and online job marketplaces, these studies primarily focus on hiring processes, overlooking the broader technological landscape that shapes domestic work. Yet, the relationship between domestic workers and technology extends beyond job searching and the hiring process; it involves ongoing coordination, task management, and social support, all of which are increasingly mediated through a range of diverse online technologies.

Our study investigates the broader landscape of online technologies employed by domestic workers, extending beyond digital labor platforms to explore technologies that support their daily tasks, work networking, and advocacy. Unlike prior studies that emphasize hiring mechanisms, we examine how workers leverage online technologies, which may include messaging applications, social media platforms, and specialized mobile applications, to coordinate their work, seek advice, and build professional resilience in environments where oversight is minimal. By doing so, we aim to provide a more comprehensive and holistic understanding of the role of technology in domestic work and highlight the agency of workers in adapting online technologies to suit their needs.

\subsection{The Digital Divide}
Previous research has delved into understanding the differences in technological adoption between individuals based on their socioeconomic and cultural factors. The ``Digital Divide'' theory has been widely used to examine disparities in access to digital technology, highlighting how these inequalities contribute to gaps in opportunities, economic outcomes, and social inclusion \cite{gordo2002planning, van2017digital,husing2006impact}. These inequalities involve the quality and availability of equipment, user autonomy, supportive social networks, and digital proficiency \cite{hargittai2003digital, van2020digital}. Different studies underscore that these socio-economic factors, such as education, income, race, gender, and age, can intensify these digital divides, limiting the benefits users derive from technology \cite{mossberger2003virtual, warschauer2004technology}.

Most studies identify three levels of digital divides between individuals and technology \cite{Warschauer2010-ds, attewell2001comment, ragnedda2017social, scheerder2017determinants}. The first level emphasizes inequalities in \textit{access} to digital technologies and infrastructure \cite{eastin2015extending, yu2006understanding}. The most common reasons include economic disparities, geographic locations, and lack of Internet in certain areas. Without access, individuals cannot develop the necessary digital skills or fully engage in opportunities available online, reinforcing broader socio-economic inequalities. Governments, companies, and institutions have aimed to reduce this divide through public policies that strengthen connectivity and democratize Internet access, such as smartphones, public WiFi initiatives, subsidies, and digital infrastructure investments \cite{middleton2010approaching,stump2008exploring,boczkowski2021abundance}. 

As the Internet has become more widespread, scholars have argued that connectivity alone does not ensure the benefits of technology. This perspective introduced the second-level digital divide, \textit{usage} \cite{dimaggio2001digital,compaine2001re,buchi2016modeling}. It considers inequalities in the quality of access and users' autonomy, often influenced by factors such as digital literacy, education, or technology proficiency \cite{hargittai2001second}. For example, Wei et al. \cite{Wei28022011} found that individuals with higher education levels and socioeconomic status are more capable of finding, interpreting, and utilizing information from the Internet. Additionally, individuals who were not exposed to online technologies early in life are more likely to feel intimidated and hesitant to explore them \cite{hill2015older,friemel2016digital}. Efforts to mitigate this divide have included improving technologies' usability and interfaces and leveraging people's digital literacy skills \cite{shneiderman2001design}.

Lastly, researchers have focused on examining the disparities in the benefits and outcomes derived from the technological divide. This third level, known as the \textit{outcome} divide \cite{wei2011conceptualizing}, explores how those who use technology more strategically or effectively can gain more significant advantages compared to those who do not use technologies, which results in more inequalities in wealth, knowledge, and opportunities \cite{helsper2012corresponding, ragnedda2017social, van2015third}. For example, Zhao and Elesh \cite{zhao2007second} argued that online technologies contribute to disparities, as only certain individuals have access to valued online networks that provide greater resources and social capital. Similarly, Hui et al. \cite{Hui2023} found that small business managers in Detroit faced several economic disadvantages by not integrating online technologies into their operations, putting them at greater risk of `being left behind' (Page 331:2). Lastly, Calder\'on-G\'omez \cite{calderon2021} argued that digital skill differences enable certain individuals to build stronger professional networks through online platforms, access specific knowledge on the Internet more efficiently, and manage broader social connections over time. As a result, the third-level divide exposes deeper structural inequalities, showing how digital literacy differences translate into real-world disparities that have been harder to address on a global scale.

\subsubsection{Immigrants and the Digital Divide}
Previous research highlights how members of disadvantaged groups, such as immigrants and low-wage workers, face disparities in digital skills and usage \cite{williams2015unified, morey2007digital}. Immigration plays a significant role in shaping digital technology use in work settings, as immigrants often face challenges in employment, cultural adaptation, health, and digital skill development \cite{elshahat2022understanding, hemminki2014immigrant, toppelberg2010language, tripp2011computer, ekoh2023understanding}. Immigrants’ engagement with technology is often driven by practical needs, such as addressing language barriers \cite{epp2017migrants}, leveraging social capital \cite{ratnam2024linking}, maintaining family communication \cite{vertovec2004cheap}, and reducing migration-related stress \cite{stafford2004maintaining}. However, barriers like language limitations, complex website interfaces, limited skills, and lack of trust hinder effective usage of digital technologies, particularly for navigating essential services like government or health insurance platforms \cite{goedhart2019just}.

At the second-level divide, trust also plays a pivotal role. Immigrants often remain on social platforms with active family and friends, limiting exposure to new technologies and online communities in the host country \cite{hsiao2023recent}. Negative perceptions of racial or ethnic differences further discourage participation in online spaces where they feel excluded \cite{van2017platform, bakan1995making}. Similarly, previous studies show that undocumented immigrants prioritize communication and convenience over privacy concerns, taking minimal precautions against surveillance risks \cite{guberek2018keeping}. Moreover, the intersectional identities of being women, Latina, and immigrants shape technology usage in unique ways. Language barriers restrict access to critical information available to English-speaking residents \cite{adkins2020information}, and many rely on basic communication tools. Immigrant parents with limited digital skills often depend on their children to navigate online resources for tasks like job searches or accessing social services \cite{katz2018connecting, rideout2016opportunity}. This dynamic, described as ``Online Search and Brokering (OSB)'' \cite{pina2018latino}, reflects how children assist with managing technology, conducting transactions, and interpreting digital content. Parents' fear of errors or security risks reinforces their reliance on family support, illustrating the persistent gaps in digital skills and usage within the second digital divide.

At the third-level divide, while digital technologies could help immigrants gain more access and information about practical resources, such as job opportunities and social connections \cite{komito2011social, lavsticova2014new, dekker2014social}, current technologies offer them limited assistance for financial, language, and cultural integration \cite{hsiao2018technology, liaqat2021intersectional}. Prior studies have shown that recent immigrants rely on social media technologies and private groups such as Facebook to establish connections with members of shared ethnicity for trusted employment resources, while more ethnically diverse groups, despite offering broader opportunities, are often avoided due to lower trust in users leading to constrained social networks and restricting opportunities for long-term integration and growth \cite{hsiao2018technology,damian2014social, hiller2004new}. Simmilarly Rodr\'iguez et al. \cite{jaimes2024costo} found that Latino house cleaners in Canada rely on digital marketplaces, such as social media commerce groups, for job opportunities, immigration support, job information exchange, mutual aid, and safety concerns. Their platform choices to access job resources are highly influenced by language, platform features, and shared identities such as ethnicity, nationality, and gender. However, this preference for shared identity platforms can have negative consequences, such as clients of the same ethnicity exploiting others through scams or low wages. Additionally, reliance on native language platforms limits job opportunities and may hinder integration into the broader society \cite{selwyn2004reconsidering, bankston2014immigrant, jaimes2024costo, hsiao2018technology}.

% Strong ties within such communities can exclude outsiders, enforce unproductive norms, or perpetuate information asymmetry by discouraging wage discussions \cite{selwyn2004reconsidering, bankston2014immigrant}. Moreover, most immigrants often trust only resources within ethnically uniform communities, leading to constrained social networks and restricting opportunities for long-term integration and growth \cite{hsiao2018technology,}. Despite these disadvantages, Rodr\'iguez et al. \cite{jaimes2024costo} found that Facebook groups provided spaces for job opportunities, immigration support, and cultural events to domestic immigrant workers in Canada, fostering solidarity and trust.

Previous work also highlights that highly skilled gig workers can maximize their incomes by using platforms' features more strategically. For example, users of the ride platform `Mystro' can obtain higher wages by filtering ride requests based on bonuses or customer ratings \cite{woodside2021bottom}. Although online platforms could provide features to help workers find better opportunities, many individuals from marginalized groups face ``information poverty,'' which limits their ability to access, interpret, and use these features properly \cite{chatman1996impoverished, chatman1991life}. Wheeler and Dillahunt \cite{wheeler2017opportunities} connected these challenges to the digital divide, emphasizing the need for equitable access to information and tools to empower workers and improve labor conditions.

\subsection{Feminist Studies and Intersectionality in HCI}
Intersectionality, a concept rooted in feminist theory, describes how race, class, gender, and other social identities intersect and overlap, shaping experiences of oppression and discrimination \cite{crenshaw2013demarginalizing}. Over time, intersectionality has been widely applied across disciplines to examine how digital inequalities emerge at the intersection of multiple identities. In HCI research, intersectionality has been used to analyze how individuals navigate disparities in access, representation, and agency in digital environments \cite{schlesinger2017intersectional}, particularly in relation to management, communication, and information systems \cite{kvasny2003triple, hedditch2023crossing}. Bardzell's \textit{Feminist HCI} agenda \cite{bardzell2010feminist} extended intersectionality into technology design, advocating for approaches that account for power dynamics, participation, and social responsibility. This agenda introduces key feminist principles, including pluralism, ecology, and advocacy, to challenge traditional user-centered design frameworks. Moreover, its concept of ecology emphasizes that technology is not neutral but deeply embedded in social, cultural, and political contexts. Overall, Feminist HCI has built upon these concepts to inform design, emphasizing that it should move beyond usability and functionality to consider long-term ethical implications and the lived experiences of marginalized users \cite{fox2017imagining, kumar2019engaging, hansson2023toolbox}. 

Enriching the concepts of feminism and intersectionality in HCI, Wong-Villacres et al. \cite{wong2018designing} advocated for shifting the focus in HCI beyond static identity categories (e.g., race or gender) to dynamic social processes. They emphasized the importance of recognizing both privileges and penalties in underserved contexts and proposed situated comparisons as a method to identify contextually relevant technology design opportunities. Additionally, the study highlighted the significance of recognizing acts of resistance, as understanding how users navigate systemic barriers can inform the design of technologies that align with diverse lived experiences. Similarly, Erete et al. \cite{erete2018intersectional} advocated for applying an intersectionality lens to examine historical oppression and discrimination, arguing that this perspective can inform equitable, inclusive, and justice-oriented design approaches for marginalized communities. Their work reinforces the need to consider the complex interplay of social, economic, and institutional factors that shape technology access, adoption, and impact.

%Transition paragraph
Building on the insights from HCI and CSCW literature, this study investigates the technological ecosystem within domestic work. Specifically, we examine how demographic, cultural, economic, and social factors influence domestic workers' adoption or avoidance of online technologies in their work practices. Understanding how they engage with specific technologies---and whether these technologies enhance or hinder their work---is important for addressing the second and third digital divides, which are strongly related to these worker communities. By centering domestic workers on technological developments, this study seeks to uncover their online practices and interactions present in domestic work, which can inform future research directions and expand current conceptualizations of technology adoption and digital divides.

\section{Methods}\label{sec:Method}
To answer our research questions, we conducted interviews with 30 house cleaners residing in the United States. All of them came from Latin America. The interviews were conducted from April to May 2024 by research team members fluent in both English and Spanish. The study was approved by the \anon{University of Notre Dame}'s Institutional Review Board (Protocol \anon{23-10-8167}).
% The study was approved by the University of Notre Dame's Institutional Review Board (IRB) under protocol number 

\subsection{Recruitment}
We recruited participants through direct outreach efforts with small cleaning service companies in four Midwestern states of the United States. The research members called managers, explained the purpose of the study, and requested them to share the research team's contact information with their workers. Workers interested in participating contacted the researchers via text message. Once the contact was established, we shared the study details with the participants and assured them that their participation would remain confidential to prevent any impact on their employment status.

All the initial messages sent to potential interviewees included a brief description of the study, expected interview length, potential risks, compensation, recording requirements, and inclusion criteria. Our inclusion criteria specified that participants must currently be working as house cleaners or have previous experience working in various settings, including private homes, offices, commercial buildings, or factories in the U.S. Additionally, participants were required to be at least 18 years old, have a minimum of three months of experience in the role, and have received payment for their services. After sharing the study information with 42 potential participants, 30 participants qualified for the interviews. From those 12 excluded individuals, some did not meet the inclusion criteria, and others withdrew their participation for personal reasons. Once a participant agreed to participate, we scheduled the interview.  

We conducted all participant interviews via phone calls to prevent the disclosure of participants' faces. At the beginning of each interview, we let participants know that the conversation was going to be recorded and ensured them of the necessary privacy, protection, and safety of their information. Thus, we emphasized that we would conduct the interview confidentially without asking for personal details such as full names, home or work addresses, or immigration status. Furthermore, we stated that no information that could potentially identify participants would be shared with others and in publications. We informed participants that their audio recordings and transcripts would be stored on our institutions' secure servers. After providing this information, participants gave their verbal consent to participate. We assigned a participant ID to each interview and removed all identifiers during the transcription and analysis process. We compensated each participant with a USD \$15 digital gift card once the data collection process ended. We sent these gift cards to their phone numbers or provided electronic mail addresses.

\subsection{Participants}
As shown in Table \ref{tab:participants}, the majority of participants identified as Latinx women ($N=29$), and most of them emigrated from M\'exico ($N=27$). Their ages ranged from 20 to 69 years, and their work experience as house cleaners varied from one to thirty years. Some participants indicated that they initially struggled to get cell phones or Wi-Fi upon their arrival in the United States ($N=23$). However, at the time of the study, none reported facing ongoing issues with accessing these technologies. All participants expressed confidence in using their cell phones for basic tasks, such as sending text messages, conducting web searches, and engaging with social platforms. Only two participants mentioned using a laptop for work-related activities. Regarding their employment status, we identified four distinct ``Employment models'' \cite{wiego2023} among our participants, which are described as follows:

\begin{itemize}
  \item \textit{Employee of Local Cleaning Companies (ELCC)}: These participants worked under the supervision of the company's management and were paid directly by the cleaning company's owner. The businesses are typically small and may be family-owned.
  \item \textit{Owner of Local Cleaning Companies (OLCC)}: Participants in this category own and operate their local cleaning businesses. Their responsibilities include acquiring clients, managing finances, and personally performing cleaning services.
  \item \textit{Direct-Hire Employee (DHE)}: These participants were hired directly by households requiring cleaning services without the intermediation of an agency or third-party company.
  \item \textit{Agency Employee (AE)}: Some participants worked for a company that acted as an intermediary between cleaning workers and clients. These agencies often operate across larger geographic areas and serve a broader client base, including both residential and commercial properties.
\end{itemize}

\begin{table}[!htb]
\small
\caption{List of study participants and relevant identifiers}
\begin{tabular}{llllcll}
\toprule
\textbf{P}   & \textbf{Age}  & \textbf{Country of origin} & \textbf{Gender} & \textbf{\makecell{Years of  \\ experience}} & \textbf{Employment model} & \textbf{Employment status} \\ 
\midrule
P1  & 40-49 &  Mexico        & Male   & 7                & OLCC     & Full-time           \\
P2  & 30-39 & Venezuela        & Female   & 7                & DHE   & Full-time             \\
P3  & 60-69 & Mexico        & Female   & 20                & AE and OLCC   & Full-time              \\
P4  & 50-59 & Mexico        & Female   & 30                & OLCC   & Full-time              \\
P5  & 50-59 & Mexico        & Female   & 20                & DHE    & Full-time             \\
P6  & 40-49 & Mexico        & Female   & 5                & AE        & Full-time         \\
P7  & 40-49 & Mexico        & Female   & 26                & DHE  and ELCC  & Full-time              \\
P8  & 50-59 & Mexico        & Female   & 20                & DHE        & Full-time         \\
P9  & 60-69 & Mexico        & Female   & 10                & DHE      & Full-time           \\
P10  & 50-59 & Mexico        & Female   & 23               & OLCC       & Full-time          \\
P11  & 40-49 & Mexico        & Female   & 15                & DHE        & Full-time         \\
P12 & 50-59 & Colombia        & Female   & 5                & DHE       & Full-time          \\
P13 & 40-49 & Mexico        & Female   & 7                & DHE         & Full-time        \\
P14 & 60-69 & Mexico        & Female   & 20               & DHE     & Full-time            \\
P15 & 20-29 & Mexico        & Female   & 4                & DHE       & Full-time          \\
P16 & 20-29 & Mexico        & Female   & 7                & DHE        & Full-time         \\
P17 & 20-29 & Mexico        & Female   & 10               & DHE      & Part-time           \\
P18 & 60-69 & Mexico        & Female   & 7                & OLCC   & Full-time              \\
P19 & 60-69 & Mexico        & Female   & 30             &  AE and DHE & Full-time               \\
P20 & 60-69 & Mexico        & Female   & 5                & DHE   & Full-time              \\
P21 & 60-69 & Honduras        & Female   & 25                & AE and DHE  & Full-time               \\
P22 & 30-39 & Mexico        & Female   & 6                & DHE         & Full-time        \\
P23 & 40-49 & Mexico        & Female   & 16                & AE and DHE   & Full-time              \\
P24 & 20-29 & Mexico        & Female   & 2                & ELCC and DHE    & Full-time           \\
P25 & 60-69 & Mexico        & Female   & 30                & ELCC    & Full-time             \\
P26 & 20-29 & Mexico        & Female   & 2                & ELCC and DHE      & Full-time           \\
P27 & 50-59 & Mexico        & Female   & 15                & ELCC and DHE & Full-time                \\
P28 & 40-49 & Mexico        & Female   & 8                & AE and DHE    & Full-time             \\
P29 & 40-49 & Mexico        & Female   & 10                & DHE     & Full-time            \\
P30 & 20-29 & Mexico        & Female   & 1                & ELCC     & Part-time           \\ \bottomrule
\end{tabular}
\captionsetup{justification=centering} % This line centers the caption
\label{tab:participants}
% including age, nationality, gender, years of experience, employment model, and employment status.
\end{table}

\subsection{Interview Protocol}
We conducted 30 semi-structured interviews via phone call, each lasting approximately 40 minutes. When asked about which language they preferred to conduct the interview, all participants opted to speak in Spanish. The first author of this paper led the interviews, while the other two co-authors joined five of the sessions to take notes, ask for clarifications, and provide feedback to the first author to enhance the quality of subsequent interviews. 

We employed a semi-structured script in our interviews to delve into participants' interactions with online technologies, their perceptions of technology adoption in their workplace, and the impact of these technologies. To guide participants through their regular activities, we structured our questions based on the ``Employment Lifecycle of Domestic Workers'' framework \cite{salih2013domestic, del2019migrant, mattingly1999job}. This framework encompasses the several processes of their work routines, from seeking employment to receiving payment. It helped the researchers and interviewees reflect on both the operational and administrative aspects of their activities and interactions with technologies. This framework consists of the following four sequential stages.

\begin{enumerate}
  \item \textit{Job search:} It covers the initial process where individuals look for available house cleaning work opportunities through various channels such as online platforms, local agencies, flyers, and personal referrals.
  \item \textit{Wage negotiation:} It involves discussions between the domestic worker and the employer regarding their compensation and activities related to wage estimation. Workers and employers discuss and agree upon the terms of compensation. This negotiation can consider hourly rates and payment for extra tasks. 
  \item \textit{Cleaning tasks:} It details the operational aspects of their work, highlighting the specific tasks and methods used in domestic cleaning. It includes a range of activities from basic tasks, such as dusting and vacuuming, to more intensive tasks like deep cleaning kitchens and bathrooms. 
  \item \textit{Payment process:} This final stage addresses domestic workers' compensation for their labor. It entails the possible payment methods, such as direct deposit, cash, or digital platforms, and the final interactions with the customer. 
  \end{enumerate}

In each one of these stages, we asked participants to what extent they employed online technologies and asked more specific questions to learn from their interactions involving online technologies and other individuals \cite{vatrapu2009towards}. To delve into the advantages and drawbacks of these interactions, we asked participants to describe a typical workday and discuss the online technologies they used. These conversations help researchers understand how domestic workers' interactions with technologies influenced their work practices and outcomes.

\subsection{Data Analysis}
We utilized Amazon Transcribe Service \cite{aws2024transcribe} to convert each interview audio into transcripts. The co-authors manually corrected the automatically-generated transcripts provided by the service, by comparing them against the audio recordings to verify their accuracy. We analyzed the transcripts directly in Spanish and only translated the specific quotes cited in this paper into English. 

After transcribing the interviews, we adopted an iterative inductive approach \cite{saldana2021coding}, allowing themes to emerge organically from the data rather than being influenced by predetermined theories or hypotheses. This grounded method ensured our findings were rooted in participants’ lived experiences. The first author conducted the initial open coding of three interviews, generating 141 codes. The second author coded two additional interviews, resulting in 112 codes, while the third author coded one interview, producing 80 codes. Subsequently, all the authors met to discuss their findings and used affinity mapping to consolidate these codes into an initial codebook containing 164 codes. Then, the first and second authors collaboratively coded half of the transcripts. Finally, all three researchers reconciled the codes with the other coder, making adjustments—such as adding or removing codes within affinity mapping—to finalize the codebook with 38 codes. This codebook was shared with all the researchers, and each one coded the same batch of interviews (10 per coder). Subsequently, the first, second, and third co-authors conducted independent analyses of the sections outlined in the codebook, which were followed by discussions on key insights. The codes were not mutually exclusive since participants' conversations could contain multiple codes.

Once all interviews were coded, the first author and last author conducted a thematic analysis to consolidate the initial codes into main themes \cite{braun2006using}. We adopted a deductive coding approach \cite{fereday2006demonstrating} to create and refine the themes. Section \ref{section:background} served as a reference point, providing a theoretical foundation for identifying and interpreting patterns in the data. The first and last authors met to refine the initial themes several times, further developing the specific definitions and boundaries of each theme. This iterative process helped the authors identify the conditions, context, and implications described by these themes. We conducted this process through several meetings, using a whiteboard to discuss themes and subthemes.

\section{Findings}
\label{sec:Findings}

Our thematic analysis identified two prominent themes salient from the interviews. The first theme, \textit{``The Bright Side: Enhancements Brought by Online Technologies,''}, explores how domestic workers adopt and adapt online technologies to their advantage, highlighting how strategically they learn, select, and employ the features of these technologies  \cite{silverstone1996design}. The second theme, \textit{``The Dark Side: Where Online Technologies Fall Short,''} examines the perceptions and underlying reasons why certain technologies are either not fully adopted or directly rejected by domestic workers. In the following subsections, we detail the distinctions between these themes and their subthemes. Table \ref{table:summary} shows a summary of these findings.

%/usepackage{multirow}
\begin{table}[!htb]
\renewcommand{\arraystretch}{1.2}
\small
\centering
\caption{Summary of the themes}
\begin{tabularx}{\textwidth}{>{\hsize=.25\hsize}X >{\hsize=.6\hsize}X >{\hsize=\hsize}X}
\toprule
\textbf{Major themes }                                  & \textbf{Sub-themes  }                              & \textbf{Definition }                                                                                                 \\ \midrule
\multirow{4}{\hsize}{\raggedright \textit{The Bright Side: Enhancements Brought by Online Technologies}} 
& Technological Mediators for Language Barriers                   & Describes how online technologies help domestic workers negotiate and establish effective communication with clients.                                           \\ \cline{2-3}
& Social Capital as a Bridge to Digital Technologies & Explores how domestic workers seek community support to expand their digital skills and utilize online technologies that enhance their work efficiency.   \\ \cline{2-3}
& Digital Strategies for Setting Prices    & Explores how domestic workers utilize online technologies to calculate and assess their services' prices.                                                         \\ \cline{2-3}
& Learning New Skills on Social Media & Examines how domestic workers employ social media platforms to enhance their skills and make their job more efficient.                                \\ \hline
\multirow{5}{\hsize}{\raggedright \textit{The Dark Side: Where Online Technologies Fall Short}}         
& The Digital Divide in Wage Comparison                          & Describes how online technologies are not utilized for comparing wages and assessing fair compensation.                              \\ \cline{2-3}
& Distrust in Online Payments                      & Participants' mistrust of clients and security concerns hinder the adoption of online payment systems.                                           \\ \cline{2-3}
& Staying Small and Invisible                         & Examines the key reasons driving domestic workers' resistance to adopt online labor platforms.  \\ \cline{2-3}
& Hopes Clicked, but Clients Did Not & Highlights that social media has not been effective in helping domestic workers find new clients or job opportunities as expected. \\ 
\bottomrule
\end{tabularx}
\label{table:summary}
\end{table}

\subsection{The Bright Side: Enhancements Brought by Online Technologies}
The interviewees described how online technologies have played a crucial role in providing them autonomy and support in some specific activities. Specifically, mobile applications on their smartphones have allowed them to leverage their work conditions and access more information in a matter of minutes. We will elaborate on the four sub-themes identified.  

\subsubsection{Technological Mediators for Language Barriers}
One of the primary challenges immigrants face is navigating language barriers. The majority of our interviewees ($N= 28$) identified themselves as basic English speakers. Despite having lived in the United States for many years, they disclosed ongoing challenges in communicating proficiently in English. This skill is crucial since job searches and wage negotiations---the first and second stages of the ``Employment Lifecycle of Domestic Worker'' framework---demand the highest levels of interaction with clients. 

To overcome these difficulties, they have developed multiple strategies to mitigate these language barriers using mobile applications. Most interviewees expressed their preference for communicating with their clients via text messages rather than in person or by phone, as this allows them to use translation applications on their smartphones effectively. These online technologies serve them as mediators, helping them clarify work arrangements and enhance communication when interacting with clients. As P23 explained: \textit{``I always prefer to move conversations with clients to text messages because I can use the translator and explain how much I will charge them and why''}. P12 added that using the translation application provided a sense of calm and control, enabling effective management of work situations: \textit{``I always tell them everything by message because I do not speak English, and that way, I can calmly use the translator. If there is any situation while I am with them in person, I use it too.''} The preference for text-based communication over phone calls or in-person discussions underscores the workers’ reliance on these online technologies to maintain control over their interactions, reduce anxiety, and achieve mutual understanding.

Interviewees, directly or indirectly, acknowledged that these challenges tend to diminish in later stages, as communication with the client becomes less frequent unless unexpected situations arise. As P7 explained: \textit{``At first, English was a problem because I had to explain and negotiate my work. But once you are hired, it is no longer necessary to speak English.''} These findings align with previous research focusing on the social dimensions of house cleaning work, which notes that house cleaners typically operate in private households and have minimal or no interaction with clients \cite{e27f23e8-5a87-37f0-a461-4deef312d79f, mattingly1999job}. 

During the ``Cleaning Tasks'' stage, interviewees reported minimal interaction with their clients. Most interviewees follow a specific cleaning routine that naturally isolates them, either because communication is unnecessary or because they prefer to avoid interactions due to their language barriers or tight schedules for completing their work. Additionally, some interviewees reported that clients who do not speak Spanish might also use translation applications to leave specific work instructions in the workspace. For example, P8 added: 
\begin{quote}
    \textit{(P8) ``I do not see my clients while I am cleaning. They are working, so I do not need to have conversations in English if there are specific instructions. They left notes in Spanish. They do not speak Spanish, but I think they use the translator to write these notes to me.''}
\end{quote}

When having interactions during this stage (e.g., a client requesting the worker to do an extra task), some interviewees used real-time online translators to ensure that they fully understood the conversation. For example, P19 said: 

\begin{quote}
    \textit{(P19) ``Most of the time, I have a routine, but if the client asks me to do some extra cleaning or gives me a specific instruction to clean something in a different way, I try to understand. But, I feel more secure when I reinforce what my client told me with the translator. I ask them to write the instructions on the translator of my cellphone.''}
\end{quote}    

Translation applications not only clarify work arrangements between workers and clients but also enhance the accuracy, clarity, and efficiency of these communications. Consequently, these online applications have enabled workers to negotiate higher wages with clients by clearly expressing the details of the cleaning tasks and the time required. Additionally, if clients expressed concerns about the price, workers could effectively explain their rates and persuade them to hire their services by detailing the different pricing options based on the type of cleaning. For example, P4, the owner of a local cleaning company, used a translator application to clearly explain the multiple cleaning services her company offered. This strategy allowed her to clearly communicate the different services to clients and specify the number of workers assigned to their houses:

\begin{quote}
    \textit{(P4) ``I have different packages of cleaning to offer to my clients, with the translator application, I explain to them what is included in each one. I do not speak English, I have been here in the US for more than 30 years, and I can not still communicate properly.''}
\end{quote}

Even during phone calls, domestic workers relied on translation applications to overcome language barriers. For instance, P18 described a practical approach, explaining how they would set their phone to translate from English to Spanish, using the microphone feature to listen and receive automatic translations: \textit{(P18) ``I give them my phone, and I set it from English to Spanish, and I listen through the microphone as it speaks and automatically translates on the phone.''}

Interviewees also reported using translators to create pre-written speeches for various scenarios, such as explaining services, requesting payments, or asking for work permissions. For example, P5 shared: \textit{``I already have a speech prepared that I always use when I need to explain the prices of my work; I used the translator to create it.''}

Finally, many interviewees expressed their frustrations with their limited English fluency, recognizing how it hindered their professional growth and interactions with clients. Despite living in the United States for many years, they voiced a strong desire to become fluent speakers but felt that their work environments offered few opportunities to practice English. The isolating nature of their jobs, combined with tight schedules and minimal client interactions, perpetuated their reliance on translation applications. As P25 acknowledged, \textit{``I have been here for 30 years and I am still not able to communicate properly. I wish that I could speak better and fluent English.''}

\subsubsection{Social Capital as a Bridge to Digital Technologies}
This subtheme highlights how domestic workers relied on their family members, coworkers, and religious groups for support in adopting and learning online technologies. They strategically used their social connections, with strong ties providing immediate support for their technological challenges and weak ties offering access to potential job opportunities. Beyond technical assistance, these social networks, particularly religious groups, also helped interviewees form a trustworthy online community that provided work-related opportunities, support, and guidance. The support network of these domestic workers primarily included daughters, sisters, female coworkers, and women from church groups, reflecting the strong role of women in these relationships.

Several interviewees described their daughters as proficient users of online technologies, often acting as brokers to support their adoption and adaptation \cite{bauer2016practising}. Their children provided assistance with a variety of activities, ranging from installing mobile applications (e.g., banking apps, translators) to teaching how to use other phone features (e.g., making notes, adding events to calendars for managing work schedules). They also guided domestic workers through more complex tasks, such as making digital banking transactions. For example, P4 described how her daughter helped her handling online technologies and explained how she relies on her to translate work-related messages from clients: 

\begin{quote}
    \textit{(P4) ``My daughter helps me a lot; she writes the messages for the clients and translates what they say for me. Moreover, she installed the translator on my cellphone. Now, I can use it when I need it.''} 
\end{quote}
 
Beyond the family network, coworkers and current clients also were fundamental to guiding domestic workers on how to use various technologies. For instance, P17 recounted how her coworkers helped her learn online technologies in her workplace, explaining, \textit{``The same people who I work with taught me how to use the translator and the dishwasher.''}. Similarly, P27 shared how a colleague taught her to operate a voice assistant device at work, stating:

\begin{quote}
\textit{(P27) ``They have on the house that we work one of these intelligent devices (Alexa), so one colleague explained to me how to put music there. Now, every time that we work, we play music on this device.''} 
\end{quote}

Church communities were also significant sources of work support for domestic workers. Weekly gatherings allowed domestic workers to network with other members, often leading to job opportunities. Additionally, church members created `WhatsApp' groups where domestic workers could access useful information about job openings. Many expressed feeling comfortable accepting and exploring opportunities shared in these online groups due to their trust in the church community. For example, P16 shared that her church was the first place that she visited when she arrived in the United States, and that was where she found her first job opportunity:
\begin{quote}
    \textit{(P16) ``In the church group, I found a lady who was looking for someone to clean her house, and that was my first job. Later, she recommended me to more people.''}
\end{quote}  

This sense of trust was further emphasized by another interviewee (P9), who explained: \textit{``When a job comes up through the WhatsApp church group, I feel confident taking it because I know they are good people and don't want to scam anyone. They always go to church, so I know them.''}

Prior research has shown that individuals who rely on their social contacts for assistance with online technology learn more quickly and gain exposure to a broader range of online services compared to those with limited social support \cite{orr2003diffusion, valente1996social, hargittai2003digital}. Our findings, reinforce the idea that social networks play a crucial role in assisting domestic workers with specific activities in the digital world. 

\subsubsection{Digital Strategies for Setting Prices}
Domestic workers reported using online technologies to assess how much they were going to charge their clients. Their price assessments were based on various factors, including time, number of bedrooms, the amount of cleaning required, the presence of pets, the type of deep cleaning needed, and the location. As an initial step, many interviewees (like P7 and P2) reported using GPS applications (e.g., Google Maps, Apple Maps, Waze) to locate and inspect properties. They will take into account factors such as distance, ZIP code, size, and the neighborhood's socioeconomic status to determine pricing:

\begin{quote}
    \textit{(P7) ``I look up the house on Google Maps, I check how big it is and the neighborhood to start figuring out how much I am going to charge them.''}
\end{quote}
\begin{quote}
    \textit{(P2) ``I always check the neighborhood on Apple Maps because if the house is near certain areas that can pay more, then I charge more. For example, if it is close to a private university, I will increase my pay because I know those students have a lot of money.''}
\end{quote}
  
However, many interviewees emphasized the limitations of these online technologies in capturing the nuances of a property, such as intricate decorations or unexpected dirtiness. As a result, many domestic workers prioritized in-person inspections for accurate price estimations. Although clients could send photos or videos, some interviewees (P18, P16, P10) reported distrust of relying solely on them, which they feared might omit critical details. For example, P18 explained that she refused to accept the job until she could visit the house and give an accurate estimation:
\begin{quote}
    \textit{(P18) ``I do not like when clients send me photos. I do not trust the clients. If I can not see the property in person, then I do not give an estimate. I need to see all the details to know how many workers I will need in each property and the time that we will invest.'' }
\end{quote} 

\subsubsection{Learning New Skills on Social Media}
Many interviewees reported using online video platforms (e.g., YouTube) and social media platforms (e.g., TikTok, Instagram, Facebook) to find tips and tutorials for new cleaning techniques. Interviewees frequently encountered cleaning tips through algorithmic recommendations. This indirect consumption of content is exemplified in the testimony of P17, who shared:
\begin{quote}
    \textit{(P17) ``I do not look for the videos directly, but when I am watching videos on TikTok or YouTube in my free time, cleaning tips come up. And then I end up watching those types of videos about how I can clean better or new devices for cleaning.''}
\end{quote}

Some interviewees demonstrated an active and direct approach to watching content related to optimizing their cleaning practices. For instance, P26 shared that YouTube helped her learn techniques to clean faster: 
\begin{quote}
    \textit{(P26) ``I search for videos on how to clean bathrooms and kitchens faster, which are generally the spaces that take the longest. I search on YouTube for tips to make them look cleaner too.''}
\end{quote}

Similarly, another domestic worker reported actively seeking information on TikTok about new chemical products and materials to enhance their cleaning work. As P29 explained: ``\textit{I really like searching on TikTok for videos about new chemicals or materials that I can use to help make the house look better. There are many new products that I can learn about thanks to TikTok.}'' 

Many domestic workers also mentioned that, despite lacking a collocated community of fellow house cleaners, they found value in joining Facebook groups and watching videos where people shared their experiences. Our interviewees reported using online communities to access and practice lateral care with other house cleaners. These online groups served as spaces for exchanging advice on handling problematic clients, as P30 explained: \textit{``I read stories from people on Facebook groups, and that way I can avoid having the same problems with clients.''} Domestic workers also used these groups to recommend products and share employment opportunities. For example, P1 shared how these online communities provided valuable insights:
\begin{quote}
    \textit{(P1) ``I really like to check out Facebook groups where there are several cleaning workers. There, I can find tips on how to grow my business, new products, and how to negotiate with clients. I belong to a Facebook group of house cleaners, and I spend some time reading their stories and experiences in this work. I do not share details about my own work, but in the future, I hope to give advice on how to run your own cleaning company.''}
\end{quote}

\subsection{The Dark Side: Where Online Technologies Fall Short}
Despite the proliferation of online technologies designed to enhance workplace efficiency, our interviewees revealed significant issues when considering or employing technologies in their daily practices. In this second theme, we describe where online technologies fail to meet domestic workers' needs. We delve into the limitations and challenges faced by workers in adopting online technologies, as well as their perceptions of the available online solutions. 

\subsubsection{The Digital Divide in Wage Comparison}
Although interviewees reported having access to multiple online resources, most of them did not use them to assess or compare their wages. Neither used the web nor other online platforms (e.g., Facebook Groups) to check if their wages were similar to the average wage. Previous research has shown that investigating and comparing wages on the Internet can lead to higher average hourly wages \cite{si2023impact, wheeler2017opportunities}. While most interviewees were uncertain about whether their compensations were fair and avoided comparing wages with other workers, only a few researched average hourly payments online.

While interviewees working as direct-hire employees and owners of local cleaning companies estimated their own wages based on their clients' properties (e.g., size of the house, the type of cleaning required) those employed by agencies could only accept wages determined by their companies or managers. Regardless of their type of employment, many interviewees were uncertain whether their earnings were fair. Many of them shared that they had never considered searching online for wage information, often citing a lack of perceived need or awareness of its potential benefits. For example, P2 reported: 
\begin{quote}
    \textit{(P2) ``To be honest, I do not know if what I earn is fair. Sometimes, I believe it is fair, but I do not have enough information to say if that is true. I have never searched on the Internet about this.''}
\end{quote}

Some interviewees were aware that their wages were low. Nevertheless, they accepted them due to fear of losing their jobs or because they received other types of compensation, such as health insurance or flexible work schedules that allow them to take care of their families. Moreover, some interviewees considered that they were old and it was risky to try to find other types of jobs with better remuneration:
\begin{quote}
    \textit{(P16) ``I have three children, and this job gives me the flexibility to pick them up from school. I know that my job is low-paid and that I could earn more doing other types of work, but as a mother, I need flexibility.'' }
\end{quote}

Two main reasons can explain interviewees' reluctance to explore or discuss their wages. First, many showed little interest in conducting online searches, preferring instead informal discussions with their peers to gather information. For example, P13 expressed that they do not need to look for this information because it may not be useful or simply because they had never considered searching for it: \textit{(P13) ``I hear this from other people doing the same job, and I base my rates on that to know if I am charging too much or too little. I have never looked it up on the internet for that reason.''}

Second, many interviewees avoid discussing economic situations as they find such conversations uncomfortable. They expressed that talking about wages can lead to tension or conflict, especially if disparities in pay among colleagues are revealed. This discomfort is often exacerbated in cultures where discussing personal finances is considered taboo or impolite. P14 exemplifies these feelings and fears: \textit{(P14) ``I do not like talking about that; it makes me uncomfortable to go around asking how much people earn. I prefer not to ask and not to be asked''}.

Furthermore, some interviewees mentioned their preference to not know about other workers' wages to avoid feelings of sadness, lack of motivation, and inferiority: \textit{(P17) ``I prefer not to know. If I find out that they earn much more than me for the same work, I would feel sad.''}. This awareness could impact their job satisfaction and overall morale, making it more challenging to approach their duties with enthusiasm and commitment. P12 expressed that the relationship with their clients could be more important than checking whether the payment was fair: \textit{(P12) ``I am very happy with my clients, and I don't want to get ideas in my head that might affect my job.''} 

Only two interviewees reported searching wage estimations on Google. By doing this, they got a better idea of how much they should charge and obtain a competitive advantage: \textit{(P28) ``I looked on Google to find out how much a domestic worker earns in my state, and I came across several websites with information. This gave me an idea of how much I should charge'' }. 

\subsubsection{Distrust in Online Payments}
The increasing use of online payment methods, such as Venmo or Zelle, has led interviewees to accept client payments through these online technologies. Many interviewees reflected on their experiences with digital payment applications, highlighting both the ease of use and the challenges associated with these online platforms. For some, learning how to manage these applications was straightforward. For instance, P14 shared, \textit{``I do not struggle to understand how to use them, but I do not like them. Actually, it was very easy to learn how to use them.''} 

Despite the simplicity of learning these online technologies, most interviewees expressed a preference for cash transactions due to concerns about trust and security. As P3 explained: \textit{``I prefer to be paid in cash. It gives me security, and I can use the money immediately.''} A recurring concern was the lack of transparency and reliability of digital payment systems. P11 voiced worries about security, stating, \textit{``I am afraid that my information and money will be stolen.''} Similarly, P3 highlighted fears of technological errors, explaining, \textit{``If I do not understand well and click on something wrong, then I am afraid that I will lose money.''} Interviewees also cited the fees associated with digital transactions as a significant drawback. As P21 noted, \textit{``They charge you extra to transfer the money, and even if it is just two dollars, you end up losing.''}

While many interviewees preferred cash payments for their immediacy and ease of access, some acknowledged the challenges of cash transactions, particularly when clients forget to pay. The lack of digital records often led to awkward conversations and delayed payments. For instance, P2 recounted an uncomfortable situation where a client disputed an unpaid amount. In contrast, P28 highlighted how such digital records were beneficial in clarifying payment discrepancies.

\begin{quote}
\textit{(P2) ``Once, a client forgot that she had paid me, and I had to tell her. She said that she had already paid me and got upset because she didn’t trust what I was telling her. It was uncomfortable.''} 
\end{quote}

\begin{quote}
\textit{(P28) ``A client told me that she had already paid me, and I told her to check her phone and verify that she had not deposited the money. It helped me a lot that everything was done over the phone; otherwise, how could I prove that she had not paid me?''}
\end{quote}

Some interviewees reported adopting online payment methods due to client preferences, despite their preference for cash. For example, P4, who manages a local cleaning company, described how transaction records facilitated accountability but still found cash payments simpler for day-to-day operations:

\begin{quote}
\textit{(P4)``If any of my employees tell me that I have not paid them, I send them a screenshot of my app or simply tell them, 'Check your account on your phone. The money is there.'... I think this is problematic and uncomfortable. For that reason, I prefer to pay all of them in cash at the end of the day. It is easier for me.''}
\end{quote}

\subsubsection{Staying Small and Invisible}
This subtheme describes how interviewees' reliance on personal networks and their skepticism toward online labor platforms. Interviewees reported that these platforms often failed to align with their job preferences. For example, many opted for consistent and recurring clients to avoid unpredictable challenges such as excessively dirty spaces or difficult clients. For example, P26 described how she carefully selected her customers:
\begin{quote}
    \textit{(P26) ``I have a specific target of clients, and I choose who to work with. I don't like arriving and finding that they have pets or that the place is very dirty.''}
\end{quote}

Interviewees expressed a strong preference for working with familiar clients to avoid risks associated with unknown individuals, such as potential mistreatment. P15 highlighted the potential risks of theft and dishonesty when hiring unknown workers through online platforms:

\begin{quote}
    \textit{(P15) ``I do not want to imagine what would happen if some of my workers stole something. It is hard to hire trustworthy and honest workers. You cannot hire through the Internet because you do not really know the person.''} 
\end{quote}

Furthermore, some interviewees expressed their satisfaction with their current workload and feared that taking on more jobs online would interfere with other responsibilities, such as family care. As P24 explained: \textit{``I do not want to take on more houses because I will not be able to cover them. If I published my services online, I would get a lot of jobs and would not be able to handle them.''} 

Domestic workers valued the trust and appreciation they received from their existing clients, which provided both security and job satisfaction. This sense of familiarity was often reinforced through personal gestures, such as inquiries about their families, holiday bonuses, or gifts. For instance, P5 emphasized how her clients valued her work and recognized her as a person, not just a worker:

\begin{quote}
    \textit{(P5) ``My clients always asked me for my family. For Mother's Day, they always give me a small gift such as chocolates. Also, in December, or when I had my daughter's XV party, some clients gave me extra money. They are really good people. I feel that they appreciate my work.''}
\end{quote}

Similarly, P8 described the kindness and personal regard shown by her clients, which motivated her to maintain these relationships rather than seek new ones through impersonal platforms:
\begin{quote}
    \textit{(P8) ``I believe that God is very good to me because I have customers who love me a lot. They congratulate me on my birthday. They ask me about my family. For these reasons, I do not want to lose them. We already have a good relationship.''}
\end{quote}

While growing their client base is appealing, finding reliable and trustworthy workers remains a significant barrier, often discouraging them from scaling their business. Instead of using online labor platforms for hiring, interviewees relied on personal networks and word-of-mouth recommendations to ensure discretion and trustworthiness, highlighting their inclination toward more cost-effective and reliable methods. Additionally, interviewees avoided online labor platforms due to high membership fees and penalties associated with cancellations: \textit{(P30) ``You have to pay for a membership, And they charge you, I think if you cannot take the job.'' }

\subsubsection{Hopes Clicked but Clients Did Not}
Although many domestic workers rely on word-of-mouth referrals to get jobs, some interviewees explored advertising strategies on social media platforms (e.g., Instagram, Facebook, TikTok) and created personal websites. Despite the amount of dedicated work, online advertisements did not help them get new clients or opportunities. For instance, P22 shared her experience combining traditional and digital advertising approaches:

\begin{quote}
\textit{(P22) ``Yes, I have tried creating my business cards and distributing them. I also have an Instagram account where I share evidence of my work to attract more clients. I make those types of publications with photos of before cleaning and after cleaning. However, I have not gotten too many clients through this method. For me, I have been more successful in getting clients thanks to their recommendations.''}
\end{quote}

Some interviewees combined online technologies with direct outreach. For instance, P1 highlighted the use of both to advertise their services:
\begin{quote}
    \textit{(P1) ``I created our own website. It was really easy because I hired a service on Google that offered many things, such as logos for the company, slogans, etc. Moreover, my wife published the before and after of our work on TikTok and Instagram. Nevertheless, most of the work that we have done has been done thanks to searching on Google for apartments, and we went there to offer our services directly.''}
\end{quote}

Overall, many interviewees noted that the most effective way to secure additional jobs was through word-of-mouth recommendations from their current clients. For example: \textit{(P7) ``In my job, I have different clients. These current clients recommend me to other clients, and that gives me more confidence to take on new clients.''} Although many interviewees considered using social media and dedicated web pages to exhibit their professional skills and advertise their services, they chose not to pursue these opportunities.  
\section{Discussion}\label{sec:Discussion}
In this study, we explored the impact of sociotechnical systems on domestic workers' activities and practices at their workplace. Through 30 semi-structured interviews, we uncovered the challenges they face and their motivations for integrating online technologies into their work. While we found that online technologies can effectively support domestic workers in making their practices and interactions with clients more effective, many of these remain inadequate for domestic workers. Drawing on previous literature and digital divide studies \cite{van2017digital, gordo2002planning, husing2006impact, van2020digital}, we discuss how these online technologies strongly influence domestic workers' practices, social relationships, and motivations. Our findings offer opportunities to inform future research and envision online technologies that align better with domestic workers’ needs, experiences, and values.

%\item \textbf{RQ1:} How do domestic workers perceive and use online technologies in their work practices, and what motivates their integration or avoidance?
First, our findings reveal that domestic workers engage with online technologies in fragmented, strategic, and context-dependent ways (RQ1). While some interviewees actively integrated online technologies into their work routines, others remained disengaged or used them in limited, selective ways, suggesting that online technology adoption is neither linear nor comprehensive across different workers. Rather, our findings suggest that technology adoption was shaped by the different job stages, employment types, family needs, work styles, and other specific demands that each worker faces. While many sociotechnical systems for work are designed around rigid, Taylorist structures that prioritize efficiency and control \cite{Khovanskaya2019,Kittur2013}, domestic workers' reported experiences reveal a far more uneven and personal process to integrate online technologies in their work practices. Even among domestic workers with similar backgrounds and experiences, integrating online technologies varied based on their specific needs, social circles, work conditions, perceived usefulness, intersectional identities, and trust in technology. Our findings underscore the complexity of technology adoption and challenge the assumption that mere access (i.e., first-level divide) or appropriate usage (i.e., second-level divide) guarantees consistent and meaningful engagement with online technologies.
%, demonstrating that adopting these technologies in their work does not follow a uniform trajectory or pattern

While the second-level digital divide is often attributed to a lack of digital skills, low socioeconomic status, and low autonomy in technology use \cite{gordo2002planning, hargittai2001second}, our findings challenge these traditional interpretations. Moreover, our study contrasts with prior research that suggests that immigrants often face challenges in developing digital proficiency and autonomy in the usage of technology \cite{goedhart_just_2019, donnellan2024capability}. In our study, domestic workers did not struggle to learn or adopt online technologies because of their backgrounds, lack of skills or understanding. On the contrary, the interviews describe how they were generally adept at using social media, translation, and GPS navigation applications. Instead, domestic workers' disengagement with technology was shaped by distrust and a lack of perceived benefit, which was particularly evident in their reluctance to integrate certain online technologies into critical work-related processes, such as online payments and job searching. As an example, many interviewees expressed concerns about losing their money or being scammed when using these online technologies. Many preferred cash payments, which provided immediate, tangible control over their earnings. These concerns align with previous research indicating that marginalized communities, particularly immigrants and low-wage workers, exhibit concerns about financial security in digital transactions \cite{pina2018latino}. Similarly, our findings are consistent with Vitak et al. \cite{Vitak2018}, who found individuals perceiving privacy and security threats online tended to avoid online technologies, opting for analog methods. Overall, our findings suggest that perceived distrust in sociotechnical systems---which includes their designers, developers, and others behind them---can reinforce the digital divide at the usage level. 

A key factor influencing domestic workers' use of online technologies was their employment model. Workers in structured environments, such as agencies or cleaning companies, had different technological needs and usage patterns than those directly hired by household owners. For example, owners of local cleaning businesses relied on social media for marketing and client acquisition, whereas direct-hire workers primarily used it for skill-building and learning cleaning techniques. Additionally, while agency employers avoided using the web to assess wages, independent workers were more open to searching for wage information on Google. Similarly, learning to use GPS applications for price assessments was more relevant for direct-hire workers, as agency workers were assigned by their managers directly. 

Furthermore, while prior research on low-wage workers suggests that online technologies are mostly employed for job-seeking \cite{10.1145/3411764.3445350, lu2021uncovering, jaimes2024costo}, we expand this literature by showing how varied and complex their usage can be. In particular, we found that many domestic workers avoid social media groups to find jobs due to trust concerns and the desire to avoid unpredictable or risky situations, leading them to search for opportunities through word-of-mouth job recommendations. This reliance on personal referrals over online technologies reflects a broader pattern in which trust-based employment structures shape digital engagement. Interestingly, WhatsApp groups within their church communities serve as a trusted medium where they feel comfortable accepting job offers, even from unfamiliar individuals. These differences highlight how each domestic worker integrates online technologies into their workflows based on their specific incentives, social contexts, and information needs. 

\subsection{Adoption Outcomes Differed Based on Domestic Workers' Motivations and Skills}
%\item \textbf{RQ2:} What are the main outcomes of integrating online technologies into the work practices of domestic workers?
As described by the third-level digital divide, our interviews reveal that some domestic workers derived more benefits from these online technologies than others, which could lead to long-term social inequalities (RQ2). For instance, some of them strategically relocated negotiation conversations to SMS to gain more control, used translation applications to interact with their clients with voice mode in real-time, investigated wage estimates on the web, or learned cleaning tips on social media. Although these online technologies were not originally designed for domestic workers, they have found novel purposes and strategic usages to meet their work needs. As a result, certain domestic workers were able to leverage these technologies to improve their outcomes, whereas others stayed behind due to trust concerns or perceived lack of utility. Participants' experiences and testimonies strengthen the persistent challenges described by the third-level digital divide, furthering deepening social inequalities among these workers \cite{wei2011conceptualizing, van2015third}. These differences in outcomes align with prior research on gig work platforms, where workers who strategically mastered platforms' features secured better wages and job opportunities than others who barely employed those features \cite{woodside2021bottom}. 

Although accessible to domestic workers and employed for certain tasks, the results suggest that online technologies have not been effectively leveraged to generate substantial social or economic capital. While these digital technologies are employed for immediate needs (e.g., communication, price assessment, task coordination), they have not facilitated activities that could meaningfully improve domestic workers' working conditions, such as higher wages or better work conditions. This limitation underscores how the adoption of online technologies remains fragmented, with their potential to build stronger economic opportunities or professional networks. Worker-centric tools that prioritize workers' benefits and well-being should be promoted in future work \cite{Zhang2022}. For example, platforms like `Dream Gigs,' can assist workers in identifying the skills they need for job opportunities \cite{10.1145/3290605.3300808}, providing strategies that are needed to reduce information poverty and enabling workers to achieve better outcomes.

We also found that domestic workers' social networks were strategically employed to obtain technological support. While family members could help with troubleshooting, colleagues could help workers find new tips or advice to improve their work.  Consistent with prior research of immigrant families \cite{pina2018latino, baron2017living, katz2016community}, we found that children and coworkers frequently act as technology brokers, assisting domestic workers in learning sociotechnical systems and bridging gaps in digital literacy. Yet, our interviewees rarely reported using online technologies to expand their current social networks. For example, many of them avoided using online labor platforms or social media platforms to find new clients and preferred to stay ``invisible.'' The lack of trust in strangers on the web made domestic workers constrained to their local networks, including church communities and immigrant communities. Our findings emphasize that trust plays a pivotal role in whether workers engage with larger communities. This reliance on immediate networks underscores the untapped potential of broader social networks to more comprehensively address digital divides, moving beyond individual assistance to systemic empowerment. To maximize the potential of social capital, online technologies should provide features that facilitate resource sharing and requests within workers' social groups. For example, wage-sharing platforms or community-driven resource hubs can encourage collaboration and knowledge exchange. %Aligning with Hsiao et al. \cite{hsiao2018technology}, we advocate for designing platforms that promote reciprocity and resource exchange within ethnic groups.

\subsection{The Crescent Moon Effect} 
Building on the second- and third-level digital divide, our findings reveal that domestic workers' engagement with online technologies is neither fully inclusive nor entirely exclusionary. Although many of the online technologies described by our interviewees were not originally designed for domestic work, they repurposed them in several ways to meet their needs, demonstrating digital adaptability rather than a lack of digital literacy or education. Many interviewees expressed confidence in adopting new online technologies, particularly when they perceived immediate, tangible benefits. This perception differs from traditional interpretations of the second-level digital divide, which often suggest that marginalized workers struggle with digital skills or autonomy in technology use because of their age or socioeconomic status \cite{hargittai2001second,katz2018connecting, rideout2016opportunity}. Instead, domestic workers engaged with technology in an intentional and selective manner, prioritizing tools that directly supported their work-related tasks. Rather than conforming to categorical classifications of technology adoption, understanding how domestic workers integrated these technologies requires examining their selective and strategic approaches, shaped by trust, perceived value, and systemic constraints. 

We describe this phenomenon as the ``Crescent Moon Effect.'' We coined this term inspired by the lunar phase in which only a small portion of the moon's surface is illuminated and visible, while the rest remains in darkness. This analogy captures the complex interplay between adoption and avoidance, illustrating how domestic workers strategically engage with certain technologies while remaining disconnected from others due to trust issues, structural barriers, or the limited perceived benefits of full digital integration. On the bright side, domestic workers successfully leverage online technologies to gain economic and social advantages in their work, such as translated conversations, skill building, and wage comparisons. However, the dark side reveals that domestic workers' engagement with certain online technologies remains incomplete, determined by fragmented skills (second-level digital divide) \cite{goedhart_just_2019} and limited economic and social outcomes (third-level digital divide). For instance, while many domestic workers used social media with ease to find cleaning tips, they avoided digital resources that could provide wage transparency resources or legal benefits information, limiting them from fully leveraging online technologies for their economic empowerment. 

This phenomenon differs from traditional digital divide frameworks, which often frame technology adoption in dichotomous terms (i.e., adopting vs. avoiding). The Crescent Moon Effect instead reveals a selective, strategic, and context-dependent interaction between domestic workers and online technologies, shaped by perceived trust, incentives, and systemic constraints \cite{Vitak2018,guberek2018keeping}. Moreover, this concept highlights the domestic workers' individual agency in selecting these online technologies \cite{bennett2023does,adam2021bridging}, repurposing certain technologies while disregarding others. Thus, the Crescent Moon Effect aims to capture the nuanced roles and motivations for technology adoption perceived by domestic workers.

\subsection{Language as a Digital Barrier and Opportunity}
As an inherent part of their immigration transitions, speaking in another language emerged as a key motivation for adopting online technologies. Although prior research has highly documented how language barriers limit integrating online technologies \cite{goedhart2019just, wheeler2018navigating}, our findings highlight the crucial role of translation applications in facilitating efficient communication at various stages of domestic work. Since many interviewees struggled to learn English, employing translation applications leveraged their ability to negotiate wages, access information, and communicate confidently with clients. As immigrants navigate the adaptation process in a new country, digital technologies are becoming essential for addressing language barriers and employment challenges \cite{hsiao2018technology, holmes2008migrants, komito2011social, dekker2014social}. 

Despite this adoption, integrating translation applications is not consistent in their work routines given the isolated nature of domestic work, where workers have limited direct interaction with clients during job performance. This lack of consistent exposure and structured learning environments prevents them from developing language skills that could improve their economic opportunities. Consistent with Liaqat's work \cite{liaqat2021intersectional} of language adaptation for immigrants, our findings reinforce that language learning opportunities for immigrants are often sporadic, unstructured, and insufficient, further exacerbating barriers to workplace communication and bargaining power. %Moreover, it is not clear how these online technologies can mitigate discrimination and biases against domestic workers.

\subsection{Domestic Work in the Digital Age}
Our work also disentangles some of the differences between domestic work and other types of labor supported by online technologies. HCI and CSCW research has distinguished the different types of labor platforms, acknowledging that those facilitating care and domestic have distinct characteristics \cite{ticona2018beyond}. While many gig-work platforms can exercise control over clients and customers through their centralized systems, real-time monitoring, and black-box algorithmic architectures \cite{rahman2023experimental}, our interviews showed the absence of central platforms in controlling domestic workers or mediating their interactions with clients. Most of the identified online labor platforms used by participants could only cover their initial transactions with their clients, but they could not fully mediate their interactions once a worker arrived at their clients' houses. Consequently, clients could exercise more power over the domestic worker, including bargaining prices or asking for extra work. Only knowing how to manage these online technologies could help domestic workers negotiate or respond better to them. By examining this triadic relationship between platforms, domestic workers, and clients \cite{Cameron2022}, researchers can better address the challenges of domestic workers, envisioning how future online technological developments and interventions can be appropriately aligned with the dynamics of this sector.

Lastly, our findings highlight a potential resistance from domestic workers to adopting online labor platforms, driven by concerns about trust, flexibility, and safety. While CSCW and HCI research has extensively examined crowd-worker and gig-work platforms like ride-sharing services \cite{seetharaman2021delivery, kusk2022platform, alvarez2023understanding, toxtli2021quantifying}, these studies often overlook emerging markets like domestic work, which demand tailored approaches. Without careful consideration, such platforms risk replicating exploitative practices common in other gig-economy sectors. Our interviewees emphasized the importance of stable, ongoing relationships with their clients, which allow them to manage additional responsibilities, such as family care, while maintaining flexibility in their schedules. They expressed caution toward digital platforms that might jeopardize these dynamics by introducing sporadic and unpredictable work schedules or failing to account for the physical demands and unexpected conditions of properties. 

\subsection{Implications and Recommendations}
Domestic workers navigate multiple intersecting identities, not only as low-wage Latina immigrant women but also as caregivers, breadwinners, and independent workers which shape their preference to avoid or adopt online technologies. Building on Wong-Villacres et al.'s \cite{wong2018designing} work, we advocate for an intersectional approach to designing online technologies that not only consider fixed identities but also analyze their dynamic social processes and the situated comparisons among domestic workers. For the situated comparison, our work reveals that the type of employment can inform distinctions of usage and outcomes. For example, agency employees typically do not engage in wage negotiations, as their salaries are pre-determined, reducing their need for negotiation-support tools. In contrast, direct-hire employees navigate complex wage discussions with clients, often in a foreign language, making translation apps and GPS-based pricing tools essential for fair negotiations. Future online technologies should facilitate context-aware digital support that aligns with diverse employment models.

Following the intersectionality framework \cite{wong2018designing}, examining how low-wage workers resist precarious work conditions can inform the development of sociotechnical systems tailored to domestic workers \cite{LowPrecariousWorkers}. For example, data-driven tools have helped workers detect wage theft \cite{WorkerCentricTools, toxtli2021quantifying} and track unpaid labor time \cite{georgiafairlabor2020wagetheft}. By analyzing these sociotechnical resistance strategies, practitioners can develop similar tools for domestic workers, enabling them to record wages, track extra hours, and document additional cleaning tasks to ensure fair compensation. As our findings demonstrate, most domestic workers do not currently use online technologies to verify their earnings, and many struggle to accurately estimate wages through online communication methods such as photos or videos. Therefore, future online technologies should focus on data-driven wage estimation and tracking systems, empowering domestic workers with tools that help them advocate for fair pay and transparency in their work arrangements. For example, Dillahunt et al. \cite{LowPrecariousWorkers} suggested using discreet smartphone surveys \cite{buskirk2012smart} to help workers identify patterns in their work conditions. Additionally, offline-friendly apps could allow workers to record their actual working hours versus agreed-upon hours to detect unpaid labor. For example, apps like Flatastic \cite{flatastic} provide a simple interface to list household tasks, track time spent on each one, and attach photos or videos of before-and-after results, helping workers collect tangible data to support fair compensation.

For our participants, the use of online technology to exchange or verify minimum wage information remains limited. While some navigate Facebook groups to compare earnings with other domestic workers, their engagement in wage-related discussions is minimal. This contrasts with the findings of Jaimes et al. \cite{jaimes2024costo}, who highlight that domestic workers in informal labor markets rely on these groups as a vital resource for understanding wage standards. Instead, our results indicate the opposite—most participants remain uncertain about their earnings, experiencing `information poverty' \cite{wheeler2017opportunities,chatman1996impoverished, chatman1991life}. This reluctance to seek wage comparisons stems from several factors. Many workers do not feel the need to verify minimum wages because they perceive other benefits based on their type of employment: agency employees can receive health insurance but fixed work schedules, whereas direct-hire employees embrace flexible schedules that allow them to balance caregiving responsibilities, at the cost of potentially lower wages. Additionally, discussing wages is often seen as uncomfortable and mentally exhausting, discouraging open exchange. Aligning with Hsiao et al. and Jaimes et al.'s recommendations \cite{hsiao2018technology,jaimes2024costo}, we advocate for the design of platforms that facilitate reciprocity and resource exchange while allowing for selective information sharing. By incorporating anonymous or controlled sharing mechanisms, domestic workers could access and compare wage information at their discretion, reducing the psychological burden of direct wage discussions while still benefiting from collective knowledge.

\subsection{Limitations and Future Work}
While our study offers valuable insights into domestic workers' usage of online technologies, it has limitations. First, our findings focus exclusively on house cleaners and may not generalize to other domestic work sectors, such as caregiving or nannying. Future research should explore diverse types of domestic work to examine unique challenges and opportunities, providing a more holistic view of the sector's technological landscape.

Second, our sample primarily consisted of Latinx domestic workers, particularly those of Mexican background. This demographic composition reflects the broader representation of Latinx workers in the U.S. domestic work sector but limits perspectives from other racial and ethnic groups, such as Black and Asian workers, who constitute 29.9\% and 6.4\% of the workforce, respectively \cite{FactSheetUSLabor}. Expanding future studies to include more diverse racial and ethnic groups, as well as workers engaged in on-demand platforms, would enhance the comprehensiveness of the findings. Stratified sampling and other recruitment strategies could address this limitation.

Third, our study did not extensively account for socioeconomic status or delve into prior interactions with technology before migration to the U.S., both common factors in the digital divide. Future work should examine these prior interactions to track how pre-migration technological experiences influence adaptation to new digital environments. Additionally, we deliberately chose not to inquire about participants’ legal immigration status to ensure their comfort and confidentiality. We acknowledge that immigration status can shape employment opportunities for domestic workers, affecting job access, stability, and reliance on digital platforms. Drawing on approaches such as \cite{guberek2018keeping}, future work should examine how undocumented status intersects with technology engagement.

Lastly, our study focused on domestic workers in the U.S., where legal status, cultural norms, and economic conditions influence technology adoption. Expanding this research to other countries would provide a broader understanding of how domestic workers engage with digital tools across different labor markets and regulatory environments.
\section{Conclusion}\label{sec:Conclusion}
Precarious work, invisible work, and low-wage work in CSCW and HCI research have explored the interactions between workers and technologies perpetuating invisibility mechanisms. Drawing on previous CSCW studies and digital divide theories, we examined how domestic workers incorporated online technologies into their practices, as well as avoided employing others. Through 30 semi-structured interviews with domestic workers residing in the United States, we learned about the social, cultural, and digital inequalities that influence their adoption of technologies at various stages of their work. We defined these integrations, obtained from the results of our thematic analysis, as the ``Crescent Moon Effect'' for technology adoption. This concept highlights the roles that online technology plays based on how domestic workers perceive and value them. Finally, we found that domestic workers were likely to avoid online technologies because of a lack of trust and low transformative change in their practices. We hope these results will provide opportunities to guide the design of more suitable online technologies, enabling an authentic live-changing role to enhance domestic workers' conditions.
%\input{08_acknowledgments}

\begin{acks}
We extend our gratitude to the house cleaners who shared their experiences with us and to the owners of small cleaning service companies who facilitated these connections. We also thank the reviewers for their constructive feedback and observations. 
\end{acks}

\bibliographystyle{ACM-Reference-Format}
\bibliography{sample-base}


\end{document}